\documentclass[preprint,superscriptaddress,showpacs,aps]{revtex4}
\usepackage{amsmath}
\usepackage{amssymb}
\usepackage{graphicx}
\usepackage{hyperref}

\newcommand{\dt}{\partial_t}
\newcommand{\vb}{{\bf v}}
\newcommand{\kb}{{\bf k}}
\newcommand{\be}{\begin{equation}}
\newcommand{\ee}{\end{equation}}
\newcommand{\bean}{\begin{eqnarray}}
\newcommand{\eean}{\end{eqnarray}}

\begin{document}

\title{Order and chaos in two-dimensional Rayleigh-B\'enard Convection} 

\author{Supriyo Paul}
\affiliation{Department of Physics, Indian Institute of Technology, Kanpur~208 016, India}

\author{Pankaj K. Mishra}
\affiliation{Department of Physics, Indian Institute of Technology, Kanpur~208 016, India}

\author{Mahendra K. Verma}
\affiliation{Department of Physics, Indian Institute of Technology, Kanpur~208 016, India}

\author{Krishna Kumar}
\affiliation{Department of Physics and Meteorology, Indian Institute of Technology, Kharagpur~721 302, India}

\date{\today} 
\begin{abstract}
A detailed study of the Rayleigh-B\'enard convection in two-dimensions with free-slip boundaries is presented. 
Pseudo-spectral method has been used to numerically solve the system for  Rayleigh number up to $3.3 \times 10^7$. The system exhibits various convective states: stationary, oscillatory, chaotic and soft-turbulent.  The `travelling rolls' instability is observed in the chaotic regime. Scaling of Nusselt number shows an exponent close to  $0.33$. Studies on energy spectrum and flux show an inverse cascade of kinetic energy and a forward cascade of entropy. This  is consistent with the shell-to-shell energy transfer in wave number space. The 
shell-to-shell energy transfer study also indicates a local energy transfer from one shell to the other.
\end{abstract}
 
 
\maketitle

\section{Introduction}

The Rayleigh-B\'enard convection (RBC) is an idealized version of the thermal convection found in nature. The study is useful for 
understanding the convective flow in atmosphere, astrophysics, crystal growth, etc. in addition to its usefulness in investigating 
the heat transport, pattern-forming instability, chaos and turbulence. In the classical RBC, a thin horizontal layer of incompressible fluid confined
between two flat plates is heated from below. The flow dynamics in RBC is  governed by two dimensionless parameters: Rayleigh number 
$R$, which is the ratio of the buoyancy and the dissipative forces, and the  Prandtl number $ P$, which is the ratio of the thermal  and the viscous diffusive 
time scales. The critical Rayleigh number $R_c$, which is the value of $R$ at the onset  of convective motion in RBC, is independent 
of $P$. The value of $R_c$ depends on the nature of boundary conditions. For thermally conducting boundaries, it is $657.5$ for 
{\it free-slip} conditions, and $1708$ for {\it no-slip} conditions on velocity fields. The Prandtl number 
$P$ affects the secondary and higher order instabilities. For non-zero $P$, the primary instability always leads to the stationary 
patterns of straight rolls~\cite{schlutter:jfm_1965}. Busse  and Clever~\cite{busse_clever:jfm_1979}  investigated extensively the 
nonlinear stability of various instabilities in  the RBC with {\it no slip} boundary conditions. Their main results are well summarized
  in the so called `Busse balloon'~\cite{Manneville:dissipative_structures}. Experiments by Krishnamurti~\cite{krishnamurti:jfm_1970a, 
  krishnamurti:jfm_1970b} showed the transition of two-dimensional (2D) convection to a three-dimensional (3D) and subsequent 
  generation of oscillatory, chaotic and turbulent convection. Ciliberto and Rubio~\cite{Ciliberto:physcripta_1987} observed in their experiments travelling waves normal to the roll axis. 

  Numerical simulations ~\cite{moore_weiss:jfm_1973,mclaughlin_orszag:jfm_1982,curry:jfm_1984,goldhirsch:jfm_1989} 
have played a significant role in the investigation of RBC. The 2D convection problem captures some of the essential features of 3D 
convection specially at higher values of $P$ ~\cite{vincent:pre_2000, schmalzl:epl_2004}. In addition, the 2D simulations are computationally less expensive compared to the 3D simulations. Curry et al.~\cite{curry:jfm_1984} investigated the 2D RBC for relatively smaller values of the reduced Rayleigh number
$r = R/R_c$  ($60 < r<  290$) and observed one and two frequency convective flows in addition to the stationary rolls. 
 Goldhirsch 
et al.~\cite{goldhirsch:jfm_1989} studied numerically the initiation of convective rolls in closed geometry. 
Scaling~\cite{biskamp:pre_63R} and turbulent behaviour~\cite{toh_matsumoto:phyfluids-2003} are also studied in 2D. 
We have simulated the convective flow in various regimes and investigated the energy spectra and fluxes.

We present here the results of  direct numerical simulations (DNS) of 2D RBC in a pure fluid ($P = 6.8$) confined between stress free 
flat boundaries. We have used various resolutions as required by the problem, and investigated systematically convective flow in 
a wide range of reduced Rayleigh number $r$ ($1.01 < r < 5\times 10^4$). We observe  ordered states (both stationary and oscillatory), 
`travelling rolls' moving chaotically in a direction normal to the roll axis, recurring ordered states,  and the soft-turbulent regime.  
The Nusselt number scales with $\epsilon = r-1$ as $\epsilon^{0.33}$. We have studied the energy spectra and fluxes for both velocity 
and temperature fluctuations. The kinetic energy shows inverse cascade, while the entropy shows forward cascade. This 
behaviour is also observed in shell-to-shell energy transfer in wave number space.  We observe a local energy transfer in the 
shell-to-shell study. 

The outline of the paper is as follows.  In Section II we describe the governing equations and the numerical method.  Section III contains descriptions of various convective states observed in our simulation.  Section IV and V contains discussions on energy spectra, fluxes, and shell-to-shell energy transfers of velocity and temperature fluctuations.  In Section VI we discuss scaling of large-scaling modes and Nusselt number with Rayleigh number.  The last section contains conclusions.   

\section{Hydrodynamic system and numerical method}

We consider a thin extended layer of Boussinesq fluid of thickness $d$, kinematic viscosity $\nu$, thermal diffusivity $\kappa$, and
thermal expansion coefficient $\alpha$ confined between two stress-free and thermally conducting horizontal plates. An adverse temperature 
gradient $\beta=\Delta{T}/d$ is imposed across the fluid layer. The hydrodynamic equations are nondimensionalized by choosing 
the length scale as $d$, velocity scale as $\kappa/d$, and temperature scale as $\Delta{T}=\beta d$. The relevant dimensionless 
hydrodynamic equations for the RBC are given by 
\bean
\dt\vb + (\vb\cdot\nabla)\vb &=& -\nabla{p}+RP\theta\hat{z}+P\nabla^2 \vb, \label{NS_eqn}\\
\dt\theta + (\vb\cdot\nabla)\theta &=& v_3 + \nabla^2 \theta, \label{heat_eqn} \\
\nabla \cdot \vb & = & 0, \label{continuity} 
\eean
where $\vb=(v_1,v_2,v_3)$  is the velocity fluctuation, $\theta$ is the perturbations in the  temperature field from the steady 
conduction state, $R=\alpha g \beta d^4/\nu \kappa$ is the Rayleigh number, and $P=\nu/\kappa$ is the Prandtl number, and $\hat{z}$ is the buoyancy direction. Two-dimensional rolls are assumed to be parallel to the $y$ axis. Free-slip and 
perfectly conducting boundary conditions at the horizontal plates imply

\bean
v_3 = \partial_{zz}v_3 = \theta = 0, ~~~~ \mbox{at}~~  z = 0, 1.  \label{bc}
\eean
The fields are considered to be periodic along the $x$ direction.

The system of equations (\ref{NS_eqn}-\ref{heat_eqn})  with the boundary conditions are numerically solved using pseudo-spectral method~\cite{Canuto}.
We use Fourier basis functions for representation  along the $x$ direction, and $\sin$ or $\cos$ functions for representation 
along $z$ direction. In this 2D simulation the velocity field is considered to be confined in the $x-z$ plane. Therefore the velocity component in the $y$-direction is considered to be zero, i.e.
\bean
	v_1 (x, z, t) &=&  \sum_{m,n} U_{m0n} (t) \exp(i m k_c x) \cos(n\pi z),\nonumber\\
	v_2 (x, z, t) &=&  0,\nonumber\\
	v_3 (x, z, t) &=&  \sum_{m,n} W_{m0n} (t) \exp(i m k_c x) \sin(n\pi z), \label{four_expansion}
\eean
where $k_c = \pi/\sqrt{2}$. Various grid resolutions, $64 \times 64$, $128\times 128$, $256\times 256$, $512\times 512$, have been used for 
the simulation.  The aspect ratio of our simulation is  $2\sqrt{2}:1$. Time stepping is carried out by the standard fourth-order 
Runge-Kutta (RK4) scheme with time steps ranging from $10^{-4}$ to $10^{-6}$.    For some runs we only excite the large-scale modes 
by properly selecting the initial conditions. The simulation is run till the system reaches  statistically steady-state.  

We carried out simulation for wide range of reduced Rayleigh number $r=R/R_c$ to explore many of the possible convective states for 2D RBC. 
We used free-slip boundary conditions on the velocity field. This condition is much convenient to formulate and use. We shall use the no-slip conditions on the velocity 
field in the future.
The range of $r$ was taken from $1.01$ to $5\times10^4$ ($R=664$ to $3.3\times 10^{7}$).  We fixed the thermal Prandtl number at $P = 6.8$ which is the typical value for 
water at room temperature.  Note that the  convection in low Prandtl number fluids is dominated by wavy rolls that is ruled out in 
2D simulation.  Therefore, we focus on large Prandtl number ($P > 1$) simulation for our 2D investigation.

\section{Simulation Results on Various Convective States}
\label{time_series_study}

Our simulation of 2D RBC spanning  a wide range of Rayleigh number exhibits a variety of convective states: stationary rolls, 
quasiperiodic rolls, chaotic flow, travelling rolls, and soft-turbulent flow. Fig.~\ref{timeseries_w11} exhibits the time series 
for large scale modes for these states  for various values of $r$.  The results in Fig.~\ref{timeseries_w11} have been observed 
for initial conditions when the imaginary parts of all the modes are nonzero, and the real parts of these modes are zero.  The 
only exceptions are  $\theta_{00n}$ modes, which are always real due to the reality condition.  When we interchange the real and 
imaginary parts of the Fourier modes (except for the  $\theta_{00n}$ modes) in the initial condition, the real and imaginary 
parts of the Fourier modes in the final states also get interchanged. 



\subsection{Ordered states}\label{ordered_states}
Figure~\ref{timeseries_w11} shows the time series for real and imaginary parts of the complex mode $W_{101}$ for various values of $r$.
 The stationary straight rolls are observed for $r \leq 80$. All the Fourier modes remain real (imaginary), if we choose them initially
  to be real (imaginary). The Fig. 1a shows the time series for stationary rolls. The stationary rolls bifurcate to the oscillatory rolls 
  for $r > 80$. All the modes still remain either real(imaginary), if the initial conditions are chosen real(imaginary). With further increase in $r$ ($r > 125$) both real and imaginary parts of all the modes develop, even if they are initially chosen as real or imaginary. 
Fig.~\ref{timeseries_w11}c shows such behavior at r=140. The real part oscillates around the zero mean and the imaginary part 
around a nonzero mean as we started simulation with imaginary values for the modes. The time period of the real part of the mode 
$W_{101}$ is twice the period of the imaginary part. For $r \geq 145$, both the real and imaginary parts of a mode show 
oscillations about nonzero mean (see Fig.~\ref{timeseries_w11}d for r=400 ). This phenomenon continues till $r = 660$. Beyond this Rayleigh number the period of real and imaginary parts develop a new frequency, 
and the system bifurcates  to a quasiperiodic state. This is observed for $660 < r < 770$ ($R = 4.3 \times 10^{5}$ to $5.1\times10^{5}$). Fig.~\ref{timeseries_w11}e shows the 
quasiperiodic time series for $r = 700$ ($R = 4.6\times 10^{5}$).

Fig.~\ref{phasespace} shows the projection of the phase space in the $\Im(w_{101})$-$\Im(\theta_{101})$ plane. The projection of 
the phase space is given for the same values of $r$ for which time-series is given in Fig.~\ref{timeseries_w11}. Notice that time 
periods of oscillations of $\Im(w_{101})$ and $\Im(\theta_{101})$  are the same for $r = 100, 140$. However, the period of 
$\Im(\theta_{101})$ is half of that for the  $\Im(w_{101})$ for $r = 400$.  All the oscillatory states observed between 
$80 < r < 660$ may be categorized in three types of different oscillatory states. Fig.~\ref{phasespace}e shows the temporal 
quasiperiodic behavior at $r = 700$; quasiperiodic behaviour is observed for $660 < r < 770$.   


\subsection{Chaos and travelling rolls}
For $770 < r < 890$ ($R= 5.1\times 10^{5}$ to $5.9\times10^{5}$), we observe that the time series of the large scale Fourier modes is chaotic (see Fig.~\ref{timeseries_w11}f).   The real and imaginary parts of a mode show small fluctuations around a mean value for long time; and suddenly they flip their sign.    Whenever the real (or the imaginary) parts of the modes change sign, the imaginary (real) parts also change sign. The imaginary and 
the real parts can also show large fluctuations without a flip. The temperature modes closely follow the  behaviour of the velocity 
modes. The projection of the phase space in $\Im(w_{101})$-$\Im(\theta_{101})$ plane is shown in Fig.~\ref{phasespace}f. An 
interesting consequence of the flip of sign of real and imaginary parts of the modes is the travelling rolls instability~\cite{MKV:arxiv_2007}. A change 
of sign in the real and imaginary parts of the  modes is equivalent to a change of phase of the complex amplitudes. The relative 
change in the phase of rolls induces a lateral movement of the 2D rolls. The convection rolls travel in a direction normal to the 
roll axis. The duration for the lateral motion is very small ($\sim 0.1$ diffusive time scale) compared to the duration for which 
the rolls do not move laterally. The `travelling rolls' instability is shown in Fig.~\ref{travelling_wave}. The three frames are 
taken just before the rolls start moving ($t=1.4$), during the motion ($t=1.43$) and just after the motion($t=1.5$).  By the time a roll travels by half of its wavelength ($\pi/k_c$), the  flow  direction reverses globally in the box~\cite{MKV:arxiv_2007}.

\subsection{Recurring ordered states}
As we increase $r$ beyond $r = 900$, the chaotic states cease to exist; instead we observe a time independent state. This stationary 
state is seen to persist for a wide range of $r$ ($900\leq r\leq 5000$). However, this state is qualitatively different from 
the time independent state observed for  $r < 80$. The spectra for both the kinetic energy and the entropy show power law behaviour which is discussed in the next section. 
 For $5000 < r \leq 10000$, we observe periodic state. The amplitude of oscillations are very 
small compared to the mean. The frequency of oscillations of the modes for this regime is very large ($\sim 4\times 10^3$). Again 
the energy spectra  $E_v (k)$ and $E_{\theta} (k)$ show power law.  The energy spectra show power law for all values of $r > 770$ 
irrespective of the temporal behaviour of the kinetic energy or entropy. Recurring ordered states (see 
Figs.~\ref{timeseries_w11}f, g and \ref{phasespace}f, g)  involve energy distribution over wide range of wave numbers with power 
law behavior.  

\subsection{Soft-turbulent state}
The soft-turbulent convection appears as $r$ is increased to $5\times 10^4$($R=3.3\times10^{7}$) (Fig.~\ref{timeseries_w11}). The structure of the flows in the soft turbulence regime agrees qualitatively with the 2D simulation for high Rayleigh number~\cite {vincent:pre_2000}.   Both the real and imaginary 
parts of the large scale velocity mode $W_{101}$ show large fluctuations around the zero mean value. Fluctuations in 
$\Re(W_{101})$ and $\Im(W_{101})$ are of the order of $200$ dimensionless units, which is much larger than the fluctuations 
observed in the $W_{101}$ mode in the periodic regime.  The probability distribution of temperature fluctuations in the middle of the box  show Gaussian 
behaviour as observed by Castaing et al.~\cite{castaing:jfm_1989} in the soft-turbulent regime. We observe breaking down of the spatial 
correlations in the flow which culminates in the turbulent motion.

\section{Energy Spectrum and Fluxes}

The kinetic energy spectrum $E_v(k)$ is defined as the sum of the kinetic 
energy  of the Fourier modes contained in the  wavenumber shell $[k,k+1)$, i.e.
\be
E_v(\kb) = \sum_{k \le k'  < k+1} \frac{1}{2}  |\vb({\bf k'})|^2.   \label{ke_spectrum}
\ee
Similarly we define the entropy as
\be
E_{\theta}(\kb) = \sum_{k \le k' < k+1} \frac{1}{2} |\theta({\bf k'})|^2.   \label{te_spectrum}
\ee

Figure~\ref{spectrum1} shows the kinetic energy and entropy spectra for relatively small Rayleigh numbers $r=70, 140, 400$.   
For this range of $r$ the kinetic energy and entropy spectrum are both exponential, i.e., $E(k) \propto \exp(-\alpha k)$.  
The exponential nature of these spectra continues up to $r=600$.   The value of $\alpha$ for kinetic energy and entropy lies 
in the range of $0.4$ to $0.8$.   This result of ours including the values of $\alpha$ are in good agreement with
Curry et.  al.~\cite{curry:jfm_1984}.  The above result indicates that only a small fraction of Fourier modes are excited appreciably for 
low values of $r$.

For $r$ beyond 600, the spectra of kinetic energy and entropy follow powerlaw behaviour as shown in Fig.~\ref{spectrum2}.   
 Note that the 2D RBC exhibits quasiperiodic ($660<r<770$), chaotic ($770<r<890$), recurring fixed points and periodic states ($900<r<10^{4}$), and 
 soft turbulent ($r$ around $5\times 10^4$) behaviour.    In the quasiperiodic, chaotic or periodic regime the power-law exponents
for the kinetic energy and the entropy are around $-4$ and $-2$ respectively.

After studying energy spectra, we turn our focus on the energy flux or energy cascade rate.  The energy flux emanating from a 
wavenumber sphere is defined as the total energy transferred from all the modes inside the sphere to all the modes 
outside the sphere~\cite{Lesieur:book}.  We compute kinetic energy flux  $\Pi_v$ and entropy flux 
$\Pi_{\theta}$ using the formalism described in Verma~\cite{MKV:physrep}.  
The flux is defined as the energy leaving per unit time from the inside of a wavenumber sphere to the outside of the sphere. This energy transfer 
takes place from the modes inside (giver) the wavenumber sphere to the modes outside (receiver) the sphere. This is computed as,
\be
\Pi_{v}(K) = \Im\left[\sum_{\kb}k_j v_i^{>}(\kb) \sum_{\bf p}v_{j}(\kb-{\bf p})v_{i}^{<}({\bf p})\right], \label{flux_compute}
\ee
where the truncated variables $\vb^{>}$ and $\vb^{<}$ are defined as follows: 
\bean
\vb^{>}(\kb) &=& \left\{ \begin{array}{c} 
0 ~~ \mbox{if} ~~ |\kb| < K,\\
\vb(\kb) ~~ \mbox{if} ~~ |\kb| > K ,
\end{array}\right . \nonumber\\
\vb^{<}({\bf p}) &=& \left\{ {\begin{array}{c} 
\vb({\bf p}) ~~ \mbox{if} ~~ |{\bf p}| < K,\\
0 ~~ \mbox{if} ~~ |{\bf p}| > K .
\end{array}}\right . 
\eean
The ${\bf p}$ summation in Eq.~\ref{flux_compute} is the convolution sum. Pseudo-spectral method can efficiently compute the flux using the truncated variables $\vb^{>}$ and $\vb^{<}$. We repeat this process for every $K$ for which we need the flux. In a similar fashion we can compute the flux $\Pi_{\theta}$ of a scalar field $\theta$.
Convective turbulence is anisotropic, yet we compute 
energy flux which is an average quantity over all the angles.  The computation of anisotropic fluxes is quite complex, and they will 
be computed in future. 

The numerically computed values for various wavenumber spheres are shown in Fig.~\ref{flux} for $r=400, 830$, and $10^4$.  
We observe negative kinetic energy flux for low wavenumbers that indicates the inverse cascade of kinetic energy.  The kinetic energy flux for a small band of higher wavenumber is positive, but its magnitude is rather small compared to the negative flux values. The inverse cascade of kinetic energy at low wavenumbers could be due to the two-dimensionality of the flow since 2D fluid turbulence exhibits inverse cascade of kinetic energy for wavenumbers lower than the forcing wavenumbers.  As shown in Fig.~\ref{flux} the energy cascade of the entropy however is forward for the full range of wavenumbers. 

 We also observe that the amount of kinetic energy flux increases with the increase of $r$.  We will discuss these issues in 
 Sec.~\ref{Sec:Large_scale}.  In the next section we will study energy transfer from a wavenumber shell to another wavenumber shell.

\section{Shell-to-shell energy transfer}

Energy flux gives the overall energy lost from a wavenumber sphere due to the nonlinear interactions.  More detailed picture of 
energy transfer is captured by another quantity called shell-to-shell energy transfer.  This quantity is specially useful for  
quantifying the locality in turbulence.  We compute the kinetic energy and entropy transferred from shell $m$ to shell $n$ 
($T^{vv}_{nm}$ and $T^{\theta \theta}_{nm}$ respectively) using the method described in Verma~\cite{MKV:physrep}.  


We define the $shell(n)$ as the shell that contains wavenumbers $shellradius(n-1) \le K < shellradius(n)$. The radii of the shells are distributed logarithmically for $4 < K < Maxpossible\_inner\_radius/2$. Hence the 
radius of the $4^{th}$ shell to the $(Nshell-3)^{th}$ shell is given by 
\be
shellradius(n) = 8\times 2^{s(n-3)}
\ee
 where $8\times 2^{s(Nshell-5)}$ $=$ $Maxpossible\_inner\_radius/2$. \\*
The wavenumber range of the shells $1$, $2$, $3$, $4$, $...$, $Nshells-1$, $Nshells$ are $[0, 2)$, $[2, 4)$, $[4, 8)$, $[8, 8\times2^s)$, $...$, $[Maxpossible\_inner\_radius/2, Maxpossible\_inner\_radius)$, $[Maxpossible\_inner\_radius, \infty)$ respectively. 
Thus, the effective shell-to-shell 
energy transfer rate from the $m$th $v$-shell to the $n$th 
$v$ -shell (Eq. (23)) can be written as ,
\be
T^{vv}_{nm} = \Im\left[\sum_{k_{n}<k<k_{n+1}}k_j v_i(\kb) \sum_{k_{m}<p<k_{m+1}}v_{j}(\kb-{\bf p})v_{i}({\bf p})\right]. \label{shell2shell_eq}
\ee
 Similarly we can compute the shell-to-shell energy transfer 
$T^{\theta\theta}_{nm}$ for the scalar field $\theta$.
 

In figure~\ref{shell2shell} we show the shell-to-shell energy transfer 
for different regimes of $r$, namely at $r=400$, $r=830$, $r=1000$, and $r=10000$. The giver shell-index $m$ is shown along y axis, 
and the receiver shell-index $n$ is shown along x axis.  In our colour scheme, red is maximum positive energy transfer, while blue 
is maximum negative energy transfer.  We can draw the following conclusions from our analysis.
\begin{enumerate}
\item Both kinetic energy and entropy shell-to-shell transfers are local, i.e., the most significant energy transfer is to the nearest 
shell, and the energy transfer to more distant shells decreases drastically.

\item For low wavenumber shells, the shell-to-shell kinetic energy transfer from $m$ to $m+1$ is negative, i.e., 
$T^{vv}_{m+1,m} <0$, while $T^{vv}_{m-1,m} = -T^{vv}_{m+1,m} > 0$.  It corresponds to inverse cascade of kinetic energy for 
low-wavenumber shells.  The sign of energy transfer for  $m \ge 5$ is reversed indicating a forward energy transfer for higher 
wavenumber shells.  These shell-to-shell energy transfer results are consistent with the flux analysis presented earlier.

\item For shell-to-shell entropy transfer, we observe that $T^{\theta\theta}_{m+1,m} > 0$ indicating forward entropy transfer.  
This result is consistent with entropy flux calculations presented earlier.

\end{enumerate}

\section{Scaling of Large Scale Modes and Nusselt Number}
\label{Sec:Large_scale}

In this Section we will describe the variation of the amplitude of large-scale modes, energy flux, and Nusselt number as a 
function of reduced Rayleigh number. Some of the large-scale velocity and temperature modes present in the system are $W_{101}$, $\theta_{101}$, and $\theta_{002}$.  In \S~\ref{time_series_study}  we discussed the dynamics of these modes at various reduced Rayleigh numbers.  Here  we describe the evolution of the amplitudes of these modes as a function of $\epsilon=r-1$.  As shown in Fig.~\ref{modes_scal},  $|W_{101}|$ grows with $\epsilon$ as $\epsilon^{0.62}$.  The amplitude of the mode $\theta_{101}$ has different scaling for small 
$\epsilon$ and large $\epsilon$.  As shown in Fig.~\ref{modes_scal}, for small $\epsilon$, $|\theta_{101}| \sim \epsilon^{0.27}$, 
but for large $\epsilon$, $|\theta_{101}| \sim \epsilon^{-0.34}$.  

We observe in our numerical simulation that $\theta_{002} \approx 0.15$ for all $r$'s.   A careful investigation of the energy equations of the large-scale mode $\theta_{101}$ nicely yields this value for $\theta_{002}$.  These arguments are described below.
the energy equation of $\theta_{101}$ is
\be
\frac{\partial}{\partial t}\frac{1}{2}\left|\theta_{101}\right|^{2}  =  T^{\theta}(101)+\Re\left[\theta_{101}^{*}w_{101}\right]
- (\pi^{2} + k_{c}^{2})\left|\theta_{101}\right|^{2},\label{eq:DNSth101}
\ee
where $k_{c} = \pi/\sqrt{2}$, and $\Re$ represents the real part. The terms  $T^{\theta}(101)$ represent the nonlinear 
interaction terms contributing to mode $\theta_{101}$. If we truncate the Fourier expansion of the velocity and temperature modes  and keep  the modes $W_{101}$, $\theta_{101}$ and $\theta_{002}$,  then the nonlinear transfer terms  turn out to be
$T^{\theta}(101) = 2\pi\Re\left[\theta_{101}^{*}w_{101}\theta_{002}\right]$. 

 We  compute various terms of  Eq. (\ref{eq:DNSth101})
using the numerical data in the 
steady-state regime.  We observe that in all regimes (ordered or chaotic) 
$T^{\theta}(101) \approx 2\pi\Re\left[\theta_{101}^{*}w_{101}\theta_{002}\right]$ within 5-10\%.  In addition to the above results, we find that the  dissipative term $(\pi^{2} + k_{c}^{2})\left|\theta_{101}\right|^{2}$ is somewhat smaller as compared to the other two 
terms. If we ignore the dissipative term,  Eq.~ (\ref{eq:DNSth101}) yields
 $\theta_{002}\approx-1/(2\pi)\approx -0.15$, which is observed quite prominently in all our DNS results.  

In addition to the above analysis of the large-scale modes, we also study the energy flux as a function of $r$.
We observe that the absolute value of maximum energy flux ($|{\Pi_v}|$) in the inverse cascade regime increases with $r$.  
In Fig.~\ref{Piu_scal} we plot  $|\Pi_v|$ as a function of $r$.  For $r$  in the range of $10^3$  to $5\times 10^4$, 
$|\Pi_v| \sim r^{1.5}$
to a good approximation.  Note that
\begin{equation}
	\Pi_v \sim u_L^3 \sim r^{3*0.6} \sim r^{1.8}
\end{equation}
where $u_L$ is the large-scale velocity.  The above estimate of $|\Pi_v|$
is in qualitative agreement with our numerical result $|\Pi_v| \sim r^{1.5}$.  The increase in the inverse energy flux with 
the increase in $r$ possibly strengthens the large-scale structure leading to fixed point or periodic solution for the large-scale 
mode $W_{101}$.   The maximum value of $|\Pi_v|$ appears to tapers off beyond $r>5\times 10^4$ where soft turbulence regime begins. 
This aspect of energy flux is under investigation. 


Another global quantity of interest is the Nusselt number which is the ratio of total heat flux and the conductive heat flux.  
 For Rayleigh number $r\leq 1$, $Nu = 1$, as only conduction takes place in this regime. Far from the threshold, Nusselt number  
 shows a power-law dependence on the Rayleigh number.  There are various predictions for the exponent some of which are described 
 below.  The theory by Malkus~\cite{malkus:proc_R_soc_1954} based on boundary layer 
stability gives an exponent of $\frac{1}{3}$. Some experimental results  \cite{Niemela_2000} are in agreement with this prediction, but other experiments
suggest  an exponent closer to $\frac{2}{7}$~\cite{castaing:jfm_1989,heslot:pra_1987,sano:pra_1989}.  It is generally believed that the Nusselt number exponents  depend on  boundary conditions and Prandtl number.  Refer to  \cite{ahlers:arxiv_2008}) for a recent review on Nuselt number scaling.

In figure~\ref{Nu_scal} we plot the Nusselt number $Nu$ as a function 
of $\epsilon = (r-1)$. 
$\epsilon$ is the measure of how far the system is from the onset. $Nu$ vs. $\epsilon$ therefore shows how the Nusselt number changes as we increase the temperature difference across the horizontal plates. 
In the oscillatory and chaotic regime we have used the mean value of $Nu$ for this plot.
The plot shows two distinct scalings for $Nu$. Close to onset, the Nusselt number increases linearly with $\epsilon$. However for $\epsilon >1$,  $Nu\sim r^{0.33\pm 0.01}$. We observe significant deviation from the fit for $100 \lesssim \epsilon \lesssim 600$;  it is interesting to note that the complex modes become time dependent around $r=125$ that lie in this band.

\section{Conclusions}
In this paper we have presented a detailed study of the 2D simulation of the RBC with free-slip boundaries for reduced Rayleigh number $r$ up to $5\times 10^5$ ($R = 3.3 \times 10^7$). The simulation is performed with a pseudo-spectral method. We observe stationary states,  time-periodic states with different features, quasiperiodic, chaotic travelling waves and soft-turbulence  for different regimes of reduced Rayleigh number $r$.   We observe that stationary and periodic states recur after chaos that indicates complex nature of the strange attractor.

We investigated the large scale complex Fourier modes $W_{101}$ and $\theta_{101}$ in these regimes. The real and imaginary parts of the complex modes couple with each other through nonlinear interaction of the higher order modes. However this coupling is active only when $r$ is raised to a sufficiently high value above $r=1$. In the chaotic regime, the real and imaginary parts of $W_{101}$ and $\theta_{101}$ show random flips. These flips lead to {\em travelling roll instability}. The rolls travel normal to the roll axis due to this instability.  This may lead to global flow reversal in the system. The results obtained are in agreement with the experimental results of Ciliberto et al.~\cite{Ciliberto:physcripta_1987}.

We analyzed energy spectra for all the convective states.  For the transition regime ($1 < r < 600$) we observe exponential energy spectra for both velocity and temperature field.  This result is consistent with that of Goldhirsch et al.~\cite{goldhirsch:jfm_1989}.   For $600 < r < 5\times 10^{4}$ within which we observe chaotic state, recurring stationary and periodic states, and soft-turbulence state, the energy spectra for both velocity and thermal fields are power law with exponents close to -4 and -2 respectively.  We have not yet investigated the energy spectra for hard-turbulence regime.   

We computed energy fluxes for various values of $r$.  We observe negative kinetic energy flux for low-wavenumber shells, and positive kinetic energy flux for intermediate wavenumber shells.  The inverse cascade of kinetic energy could be due to the two-dimensionality of the system.  This claim is not conclusive since the kinetic energy spectrum exponent is close to -4, not -5/3 as observed for the 2D fluid turbulence in the inverse cascade regime.  It is possible that  the hard-turbulence regime has -5/3 kinetic energy spectral index for low wavenumbers  (for $r>5\times 10^{4}$).   The  entropy flux is positive for all $r$ indicating a forward entropy $|\theta(k)|^2$ cascade.  

To investigate energy transfers in convective turbulence in detail we also study the shell-to-shell energy transfers for both velocity and temperature fields.  We observe the shell-to-shell transfer to be local for both the fields.  For kinetic energy, the transfer is backward for smaller wavenumber shells, and forward for the higher wavenumber shells.  However the transfer for the entropy is forward.  These results are consistent with the flux results. 

We analyzed the variation of large-scale modes  $W_{101}$, $\theta_{101}$, and $\theta_{002}$ as a function of $r$.  We 
find that $W_{101} \sim r^{0.62}$, while $\theta_{101} \sim r^{0.27}$ for small $r$ and $\theta_{101} \sim r^{-0.34}$ for large $r$.  We observe that $\theta_{002} \approx 0.15$ for all $r$.  This result has been derived using a numerical input from our simulation.  We compute the Nusselt number scaling, and find that it scales as $r^{0.33}$.

\begin{acknowledgments}
We thank Stephan Fauve, K. R. Sreenivasan, Joe Niemela, Daniele Carati, Arul Lakshminarayan, Pankaj Wahi, and Pinaki Pal discussions and  various important suggestions.  We thank Computational Research Laboratory for providing computational resource to complete this work. Part of the work was supported by the grant of Swarnajayanti fellowship by Department of Science and Technology, India.  Part of this work was done as the doctoral thesis work of SP.
\end{acknowledgments}

\newpage
\begin{center}
{\bf Figure captions}\\
\end{center}
\noindent
FIG. 1: Time series for the real and imaginary parts of the mode $W_{101}$ for  $r=70$ { (a)},  $r=100${ (b)}, 
 $r=140${ (c)},  $r=400${ (d)},  $r=700${ (e)},  $r=830${ (f)}, 
 $r=1500${ (g)},  $r=7000${ (h)},  $r=50000${ (i)}. Prandtl number is fixed at $P=6.8$. In (h) we plot $\Re(w_{101})/78.5$ and
 $\Im(w_{101})/20$ for visual clarity. \\
 
\begin{figure}
\begin{center}
\includegraphics[height=!,width=16cm]{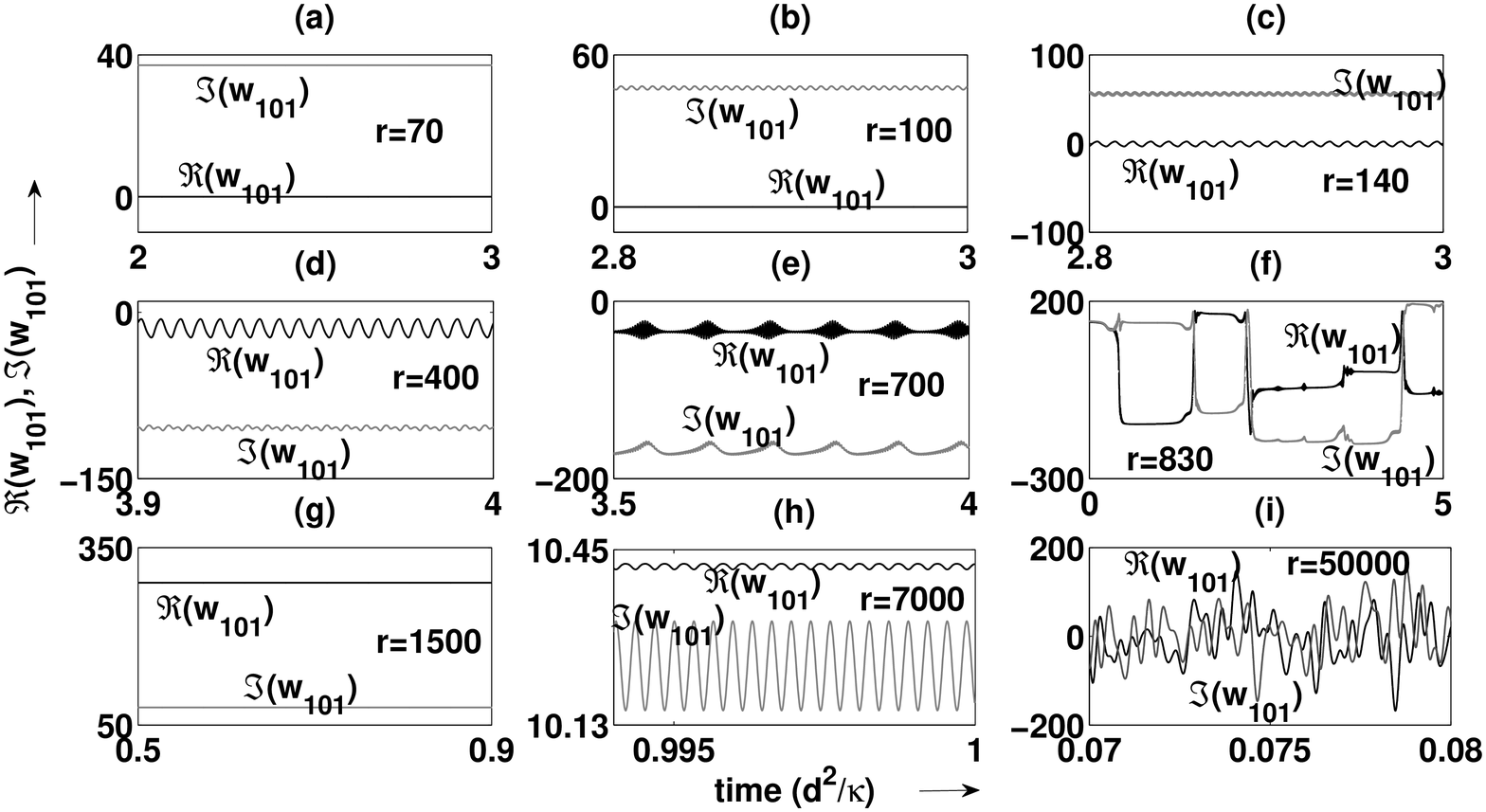}
\end{center}
\caption{}
\label{timeseries_w11}
\end{figure}

\begin{figure}
\begin{center}
\includegraphics[height=!,width=18cm]{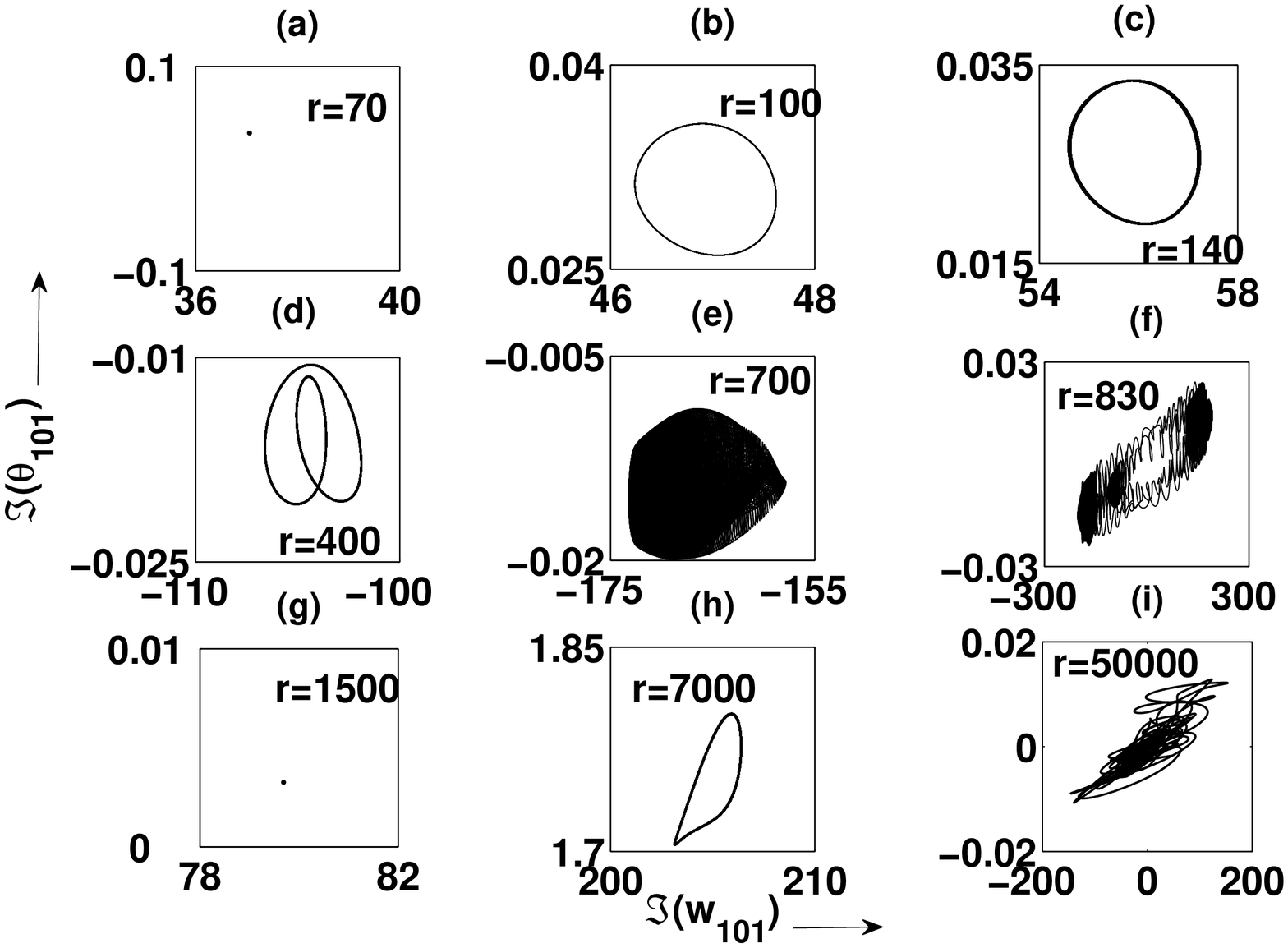}
\end{center}
\caption{}
\label{phasespace}
\end{figure}

\begin{figure}
\begin{center}
\includegraphics[height=!,width=16cm]{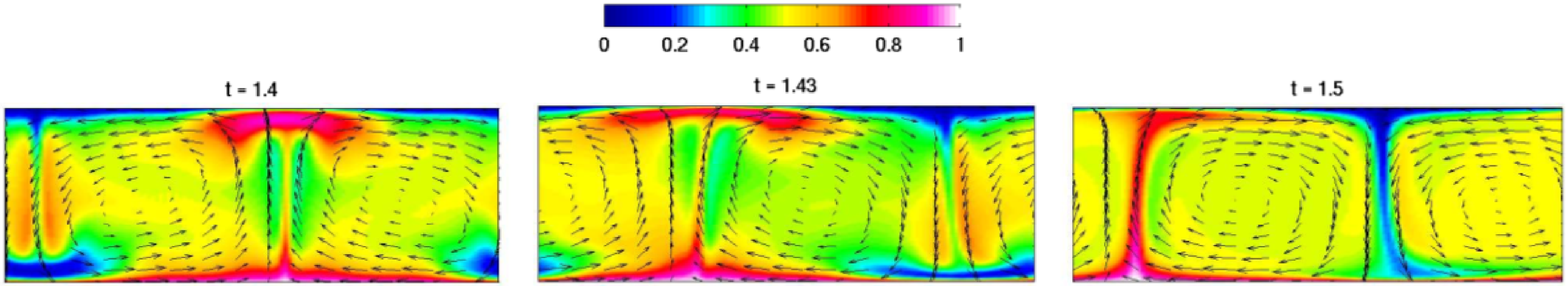}
\end{center}
\caption{}
\label{travelling_wave}
\end{figure}

\begin{figure}
\begin{center}
\includegraphics[height=!,width=14cm]{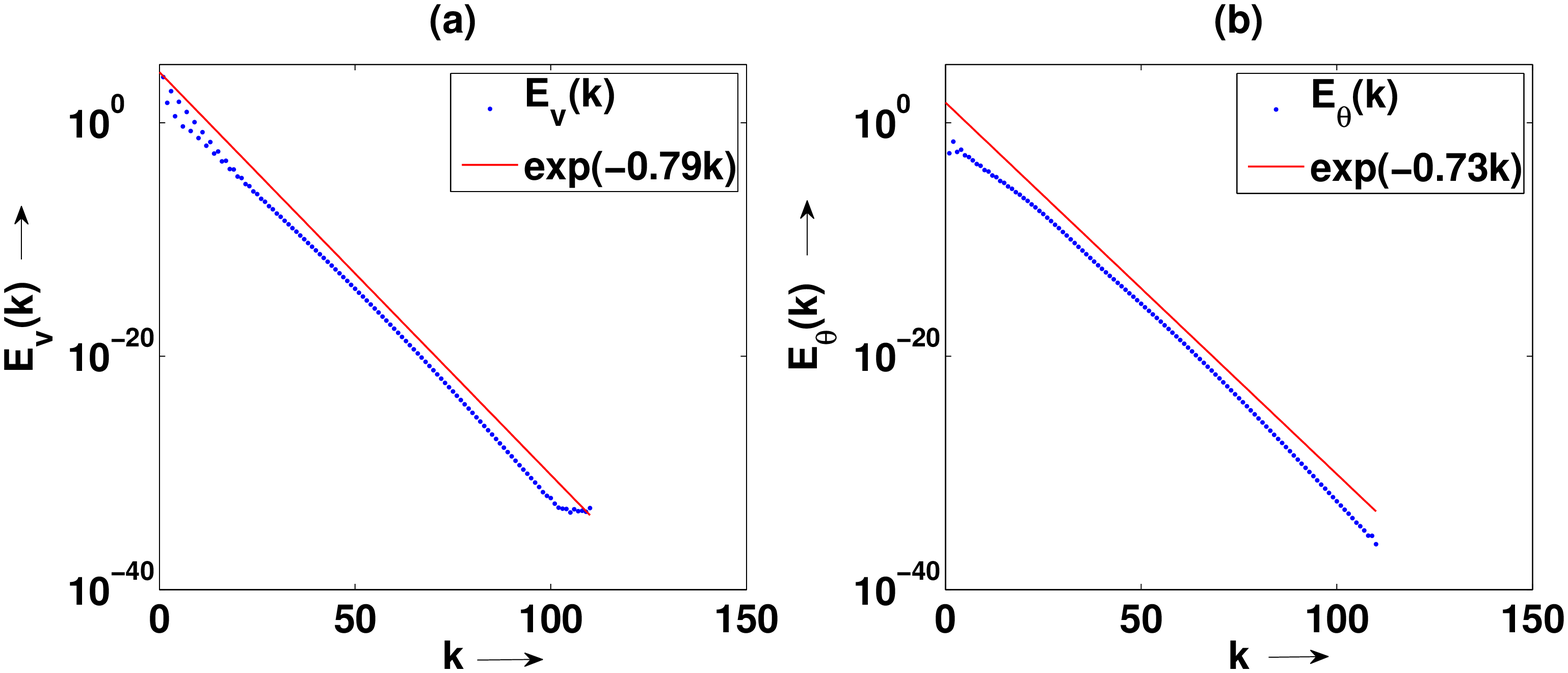}
\includegraphics[height=!,width=14cm]{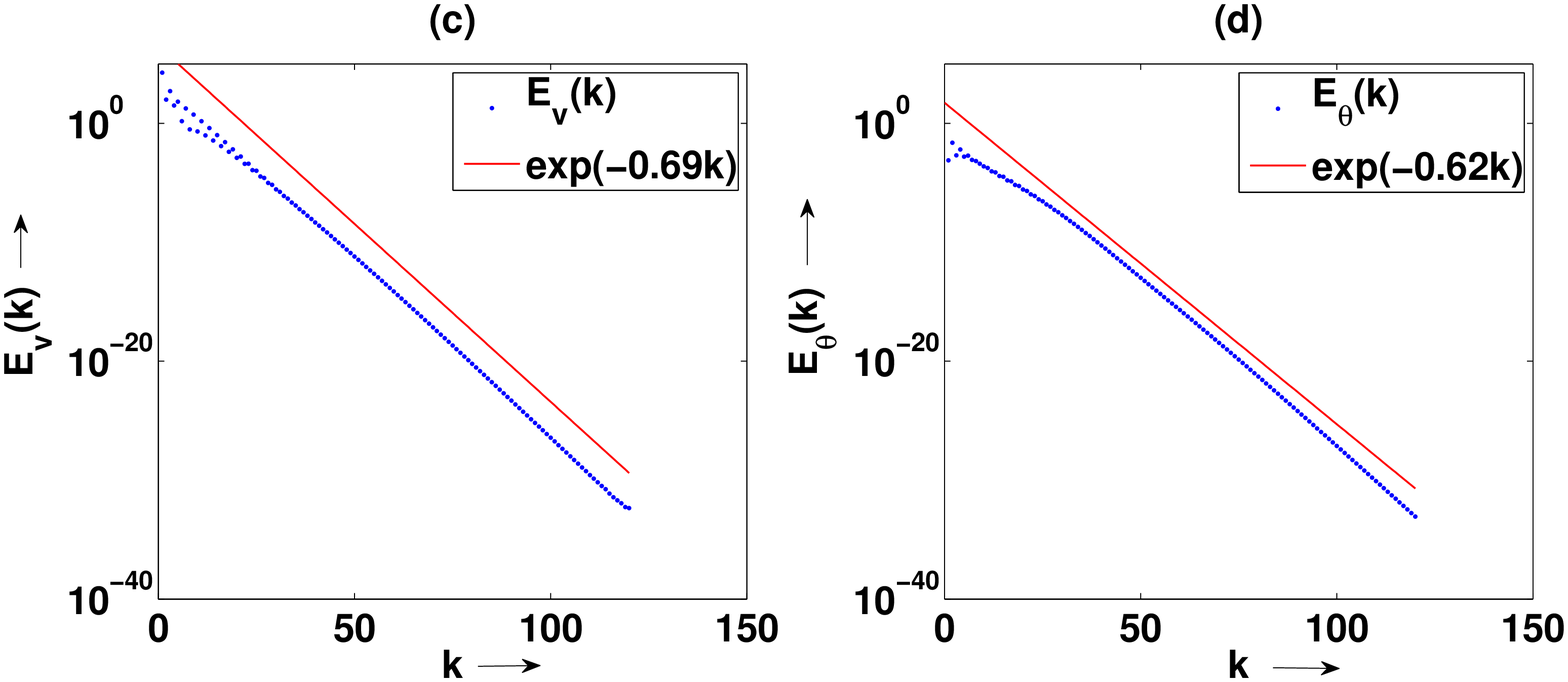}
\includegraphics[height=!,width=14cm]{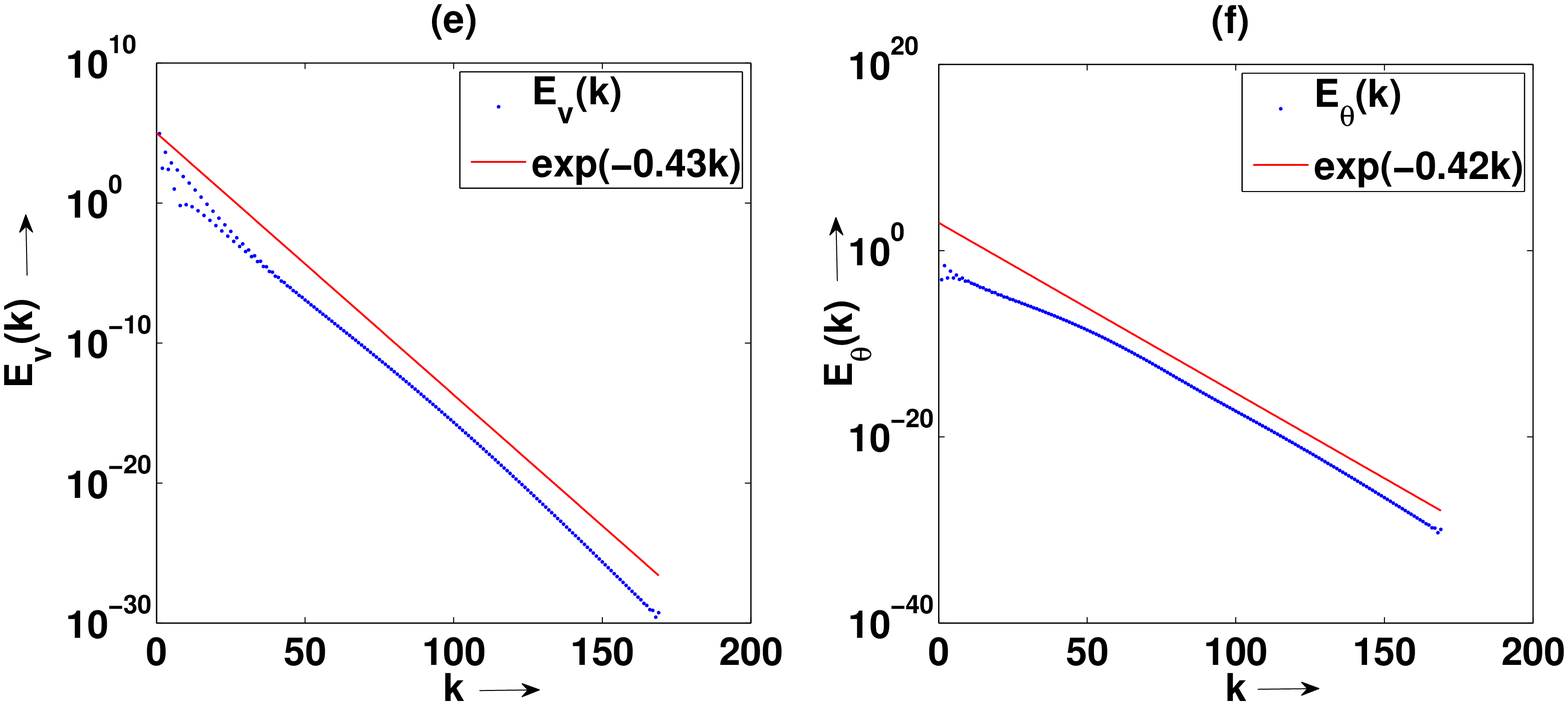}
\end{center}
\caption{}
\label{spectrum1}
\end{figure}

\begin{figure}
\begin{center}
\includegraphics[height=!,width=14cm]{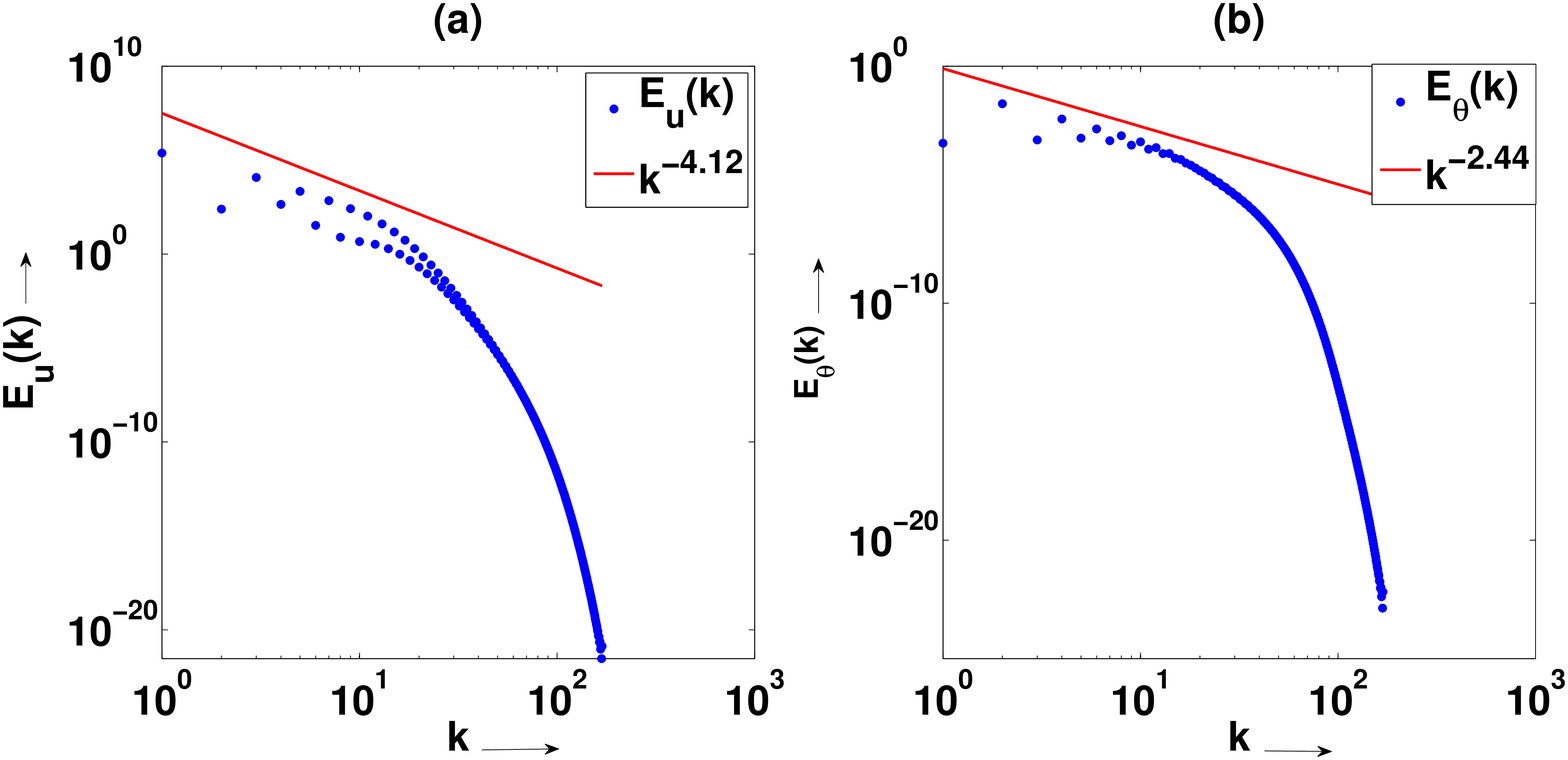}
\includegraphics[height=!,width=14cm]{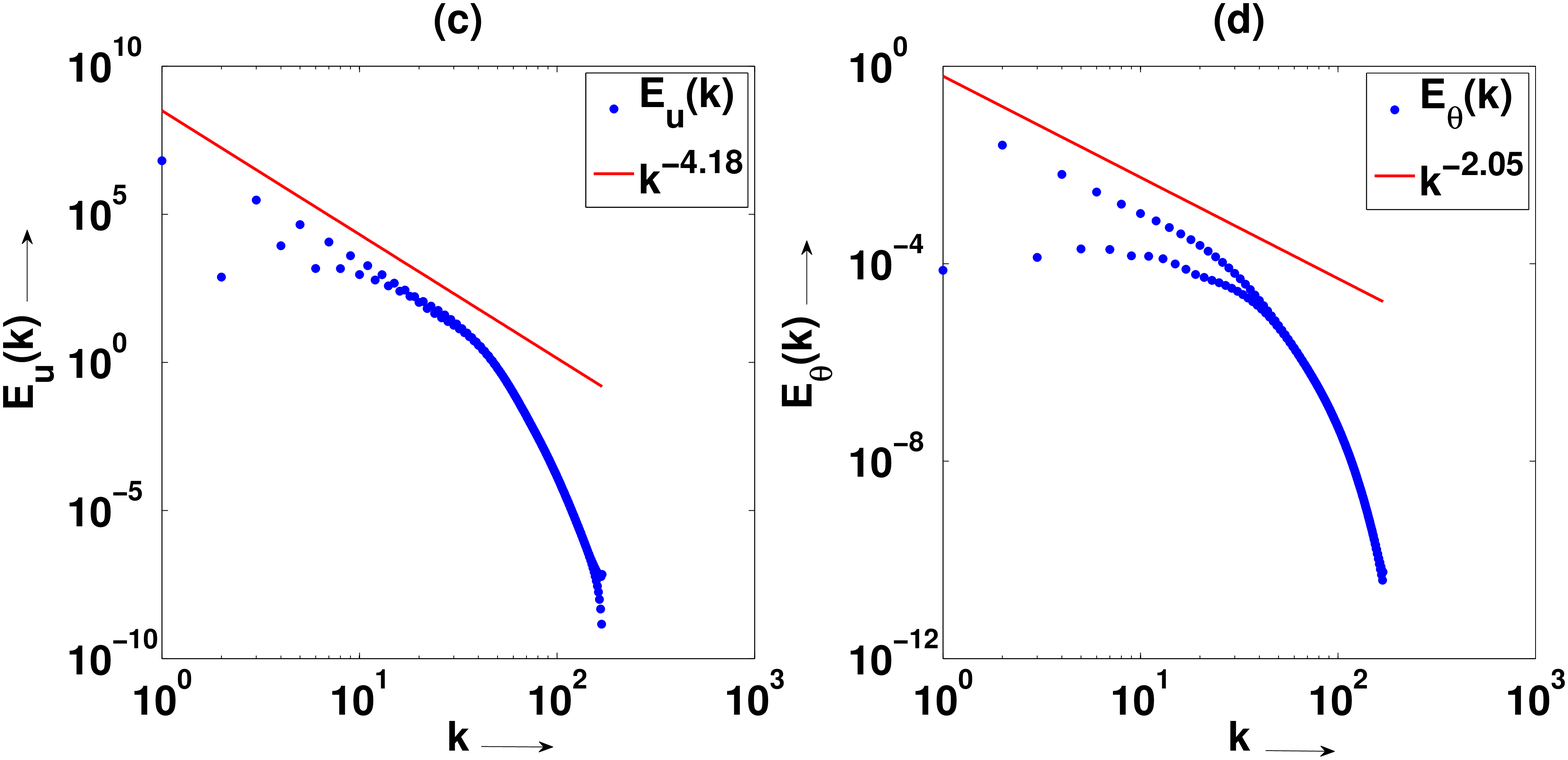}
\includegraphics[height=!,width=14cm]{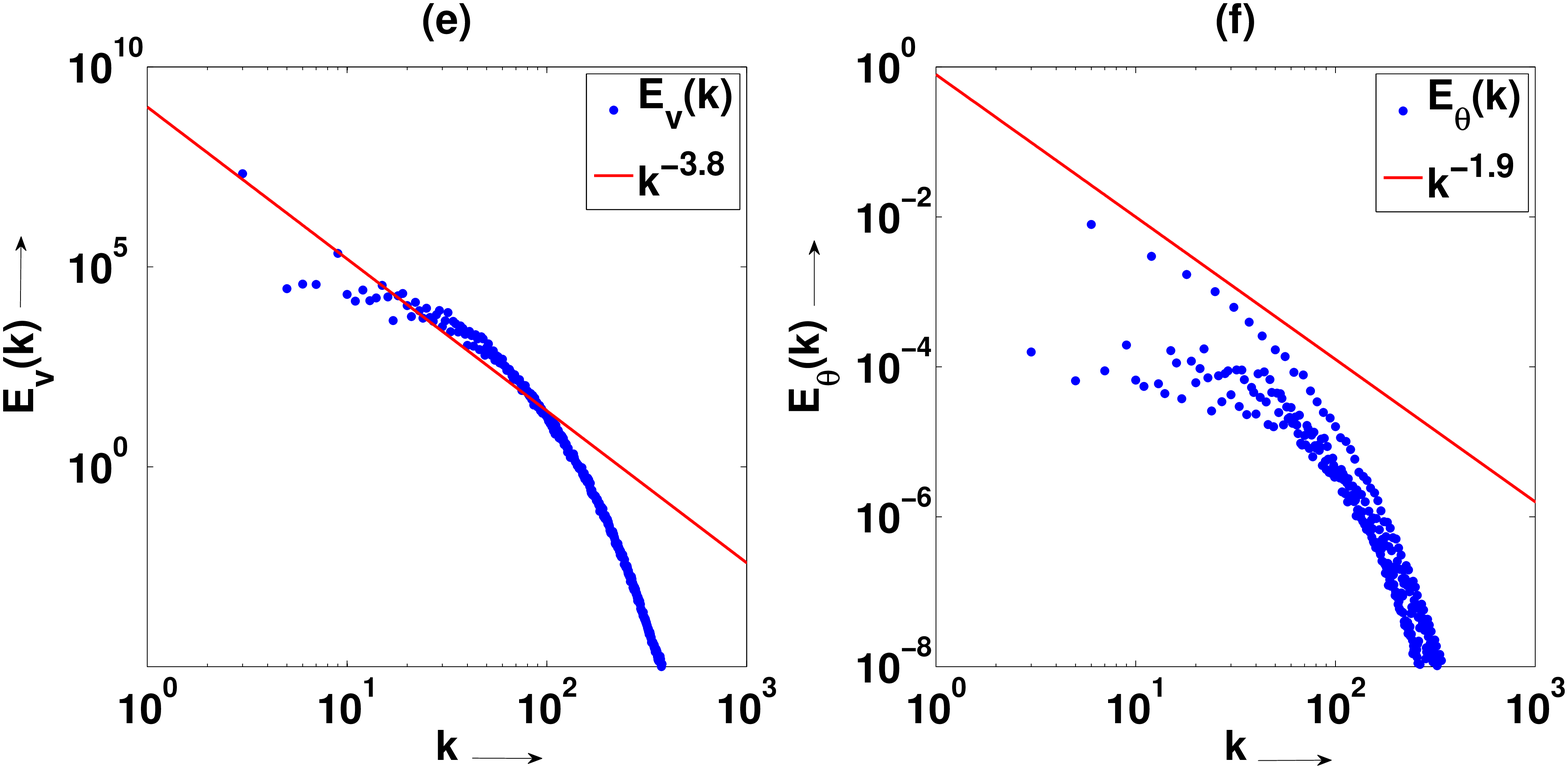}
\end{center}
\caption{}
\label{spectrum2}
\end{figure}

\begin{figure}
\begin{center}
\includegraphics[height=!,width=14cm]{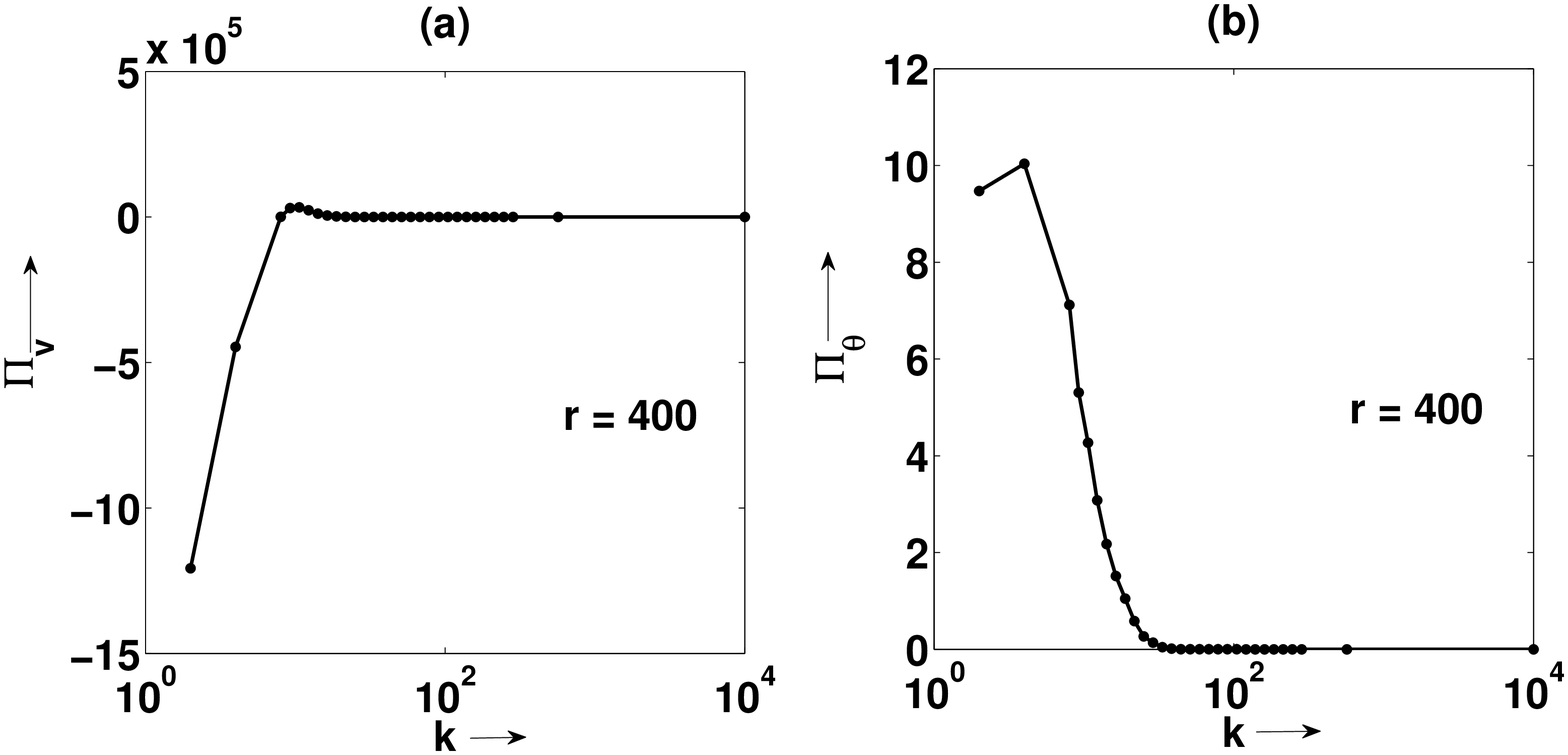}
\includegraphics[height=!,width=14cm]{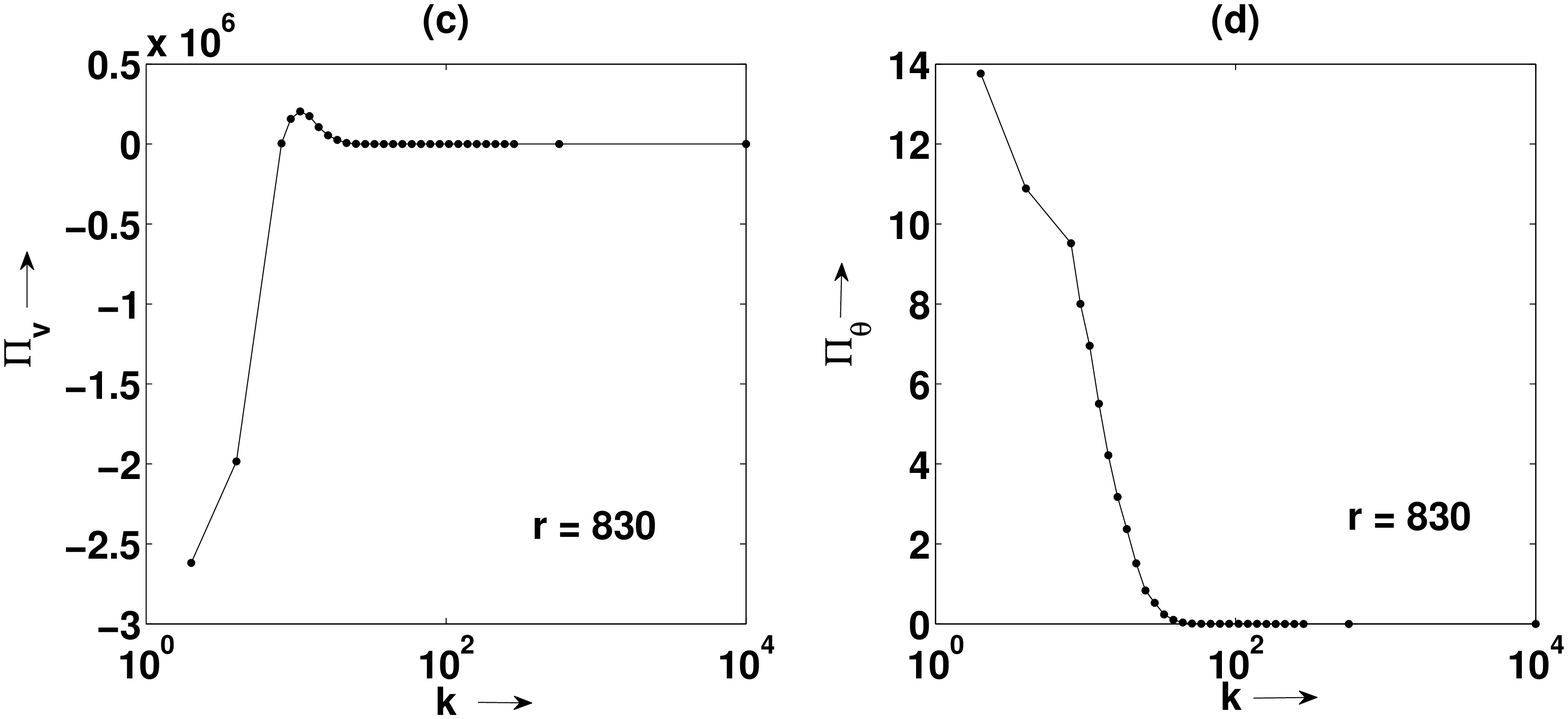}
\includegraphics[height=!,width=14cm]{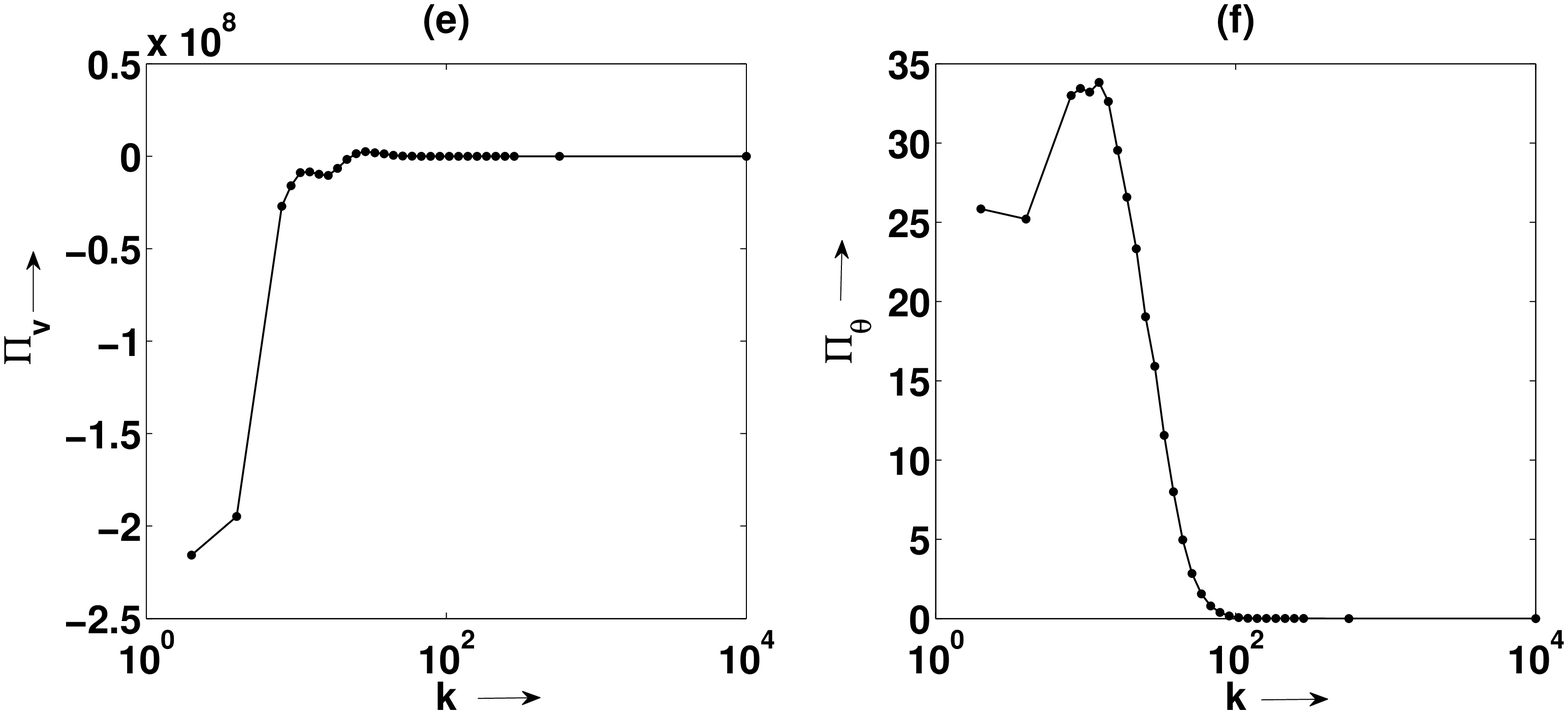}
\end{center}
\caption{}
\label{flux} 
\end{figure}

\begin{figure}
\begin{center}
\includegraphics[height=!,width=16cm]{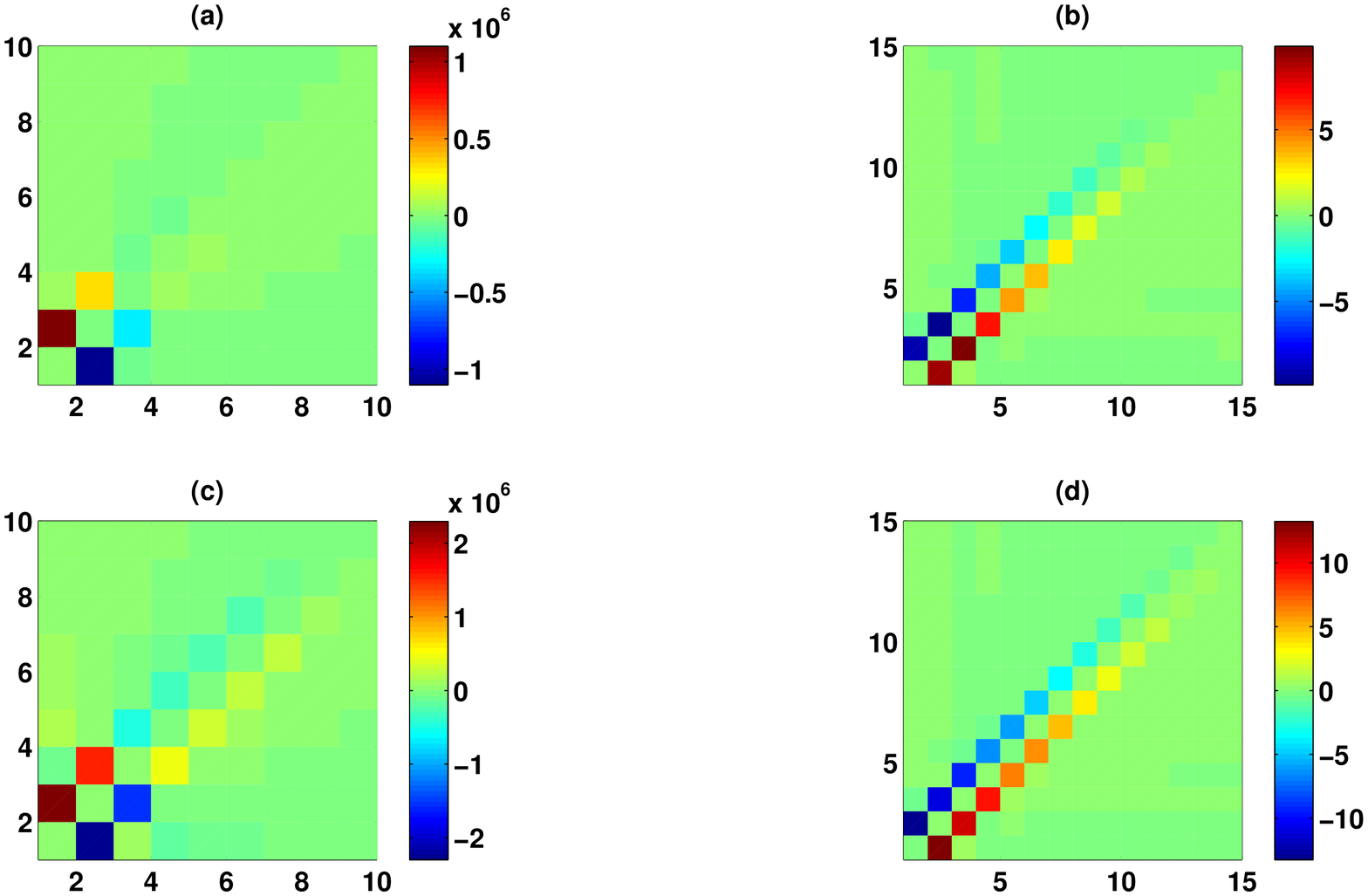}
\includegraphics[height=!,width=16cm]{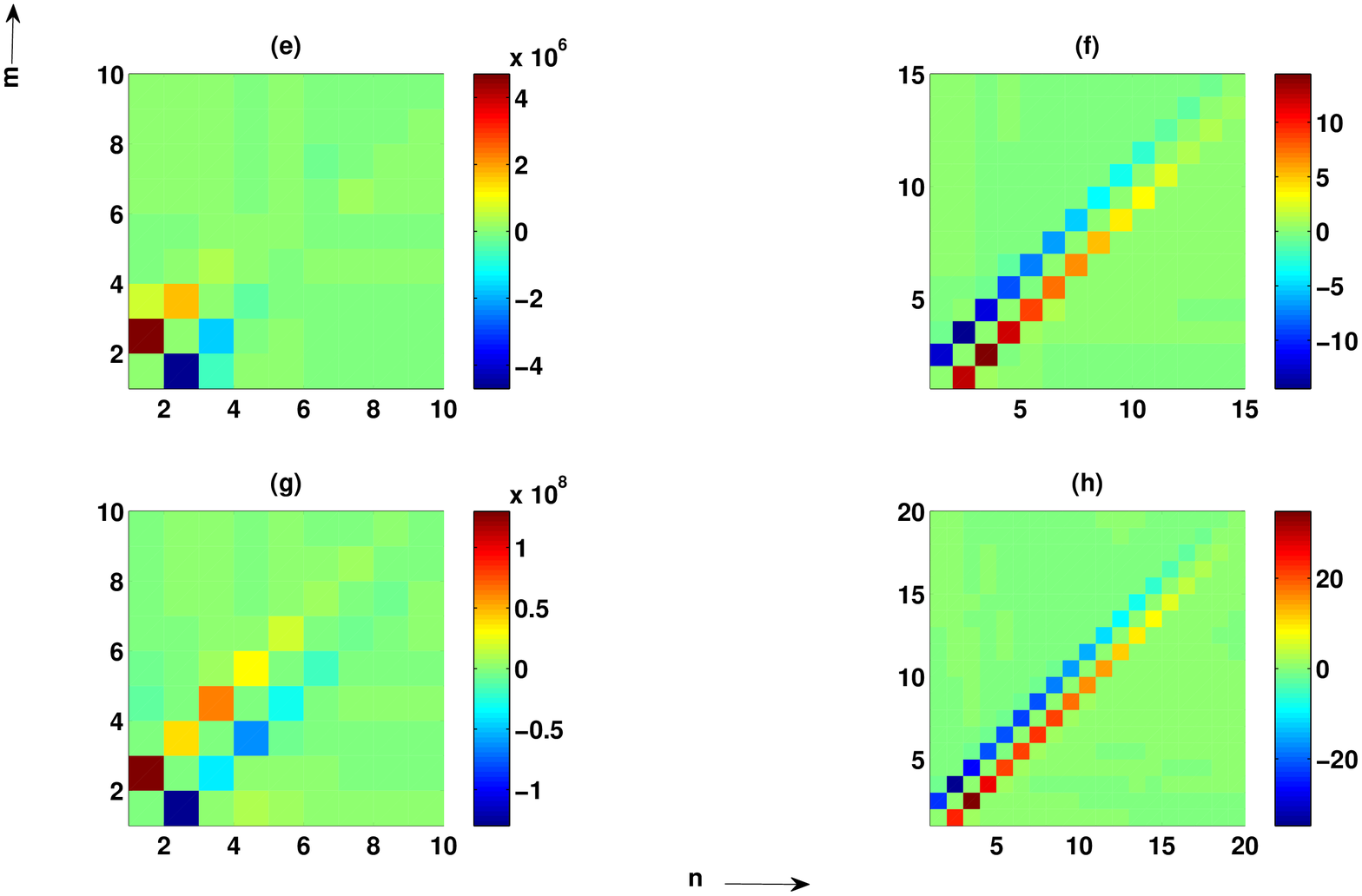}
\end{center}
\caption{}
\label{shell2shell}
\end{figure}

\begin{figure}
\begin{center}
\includegraphics[height=!,width=!]{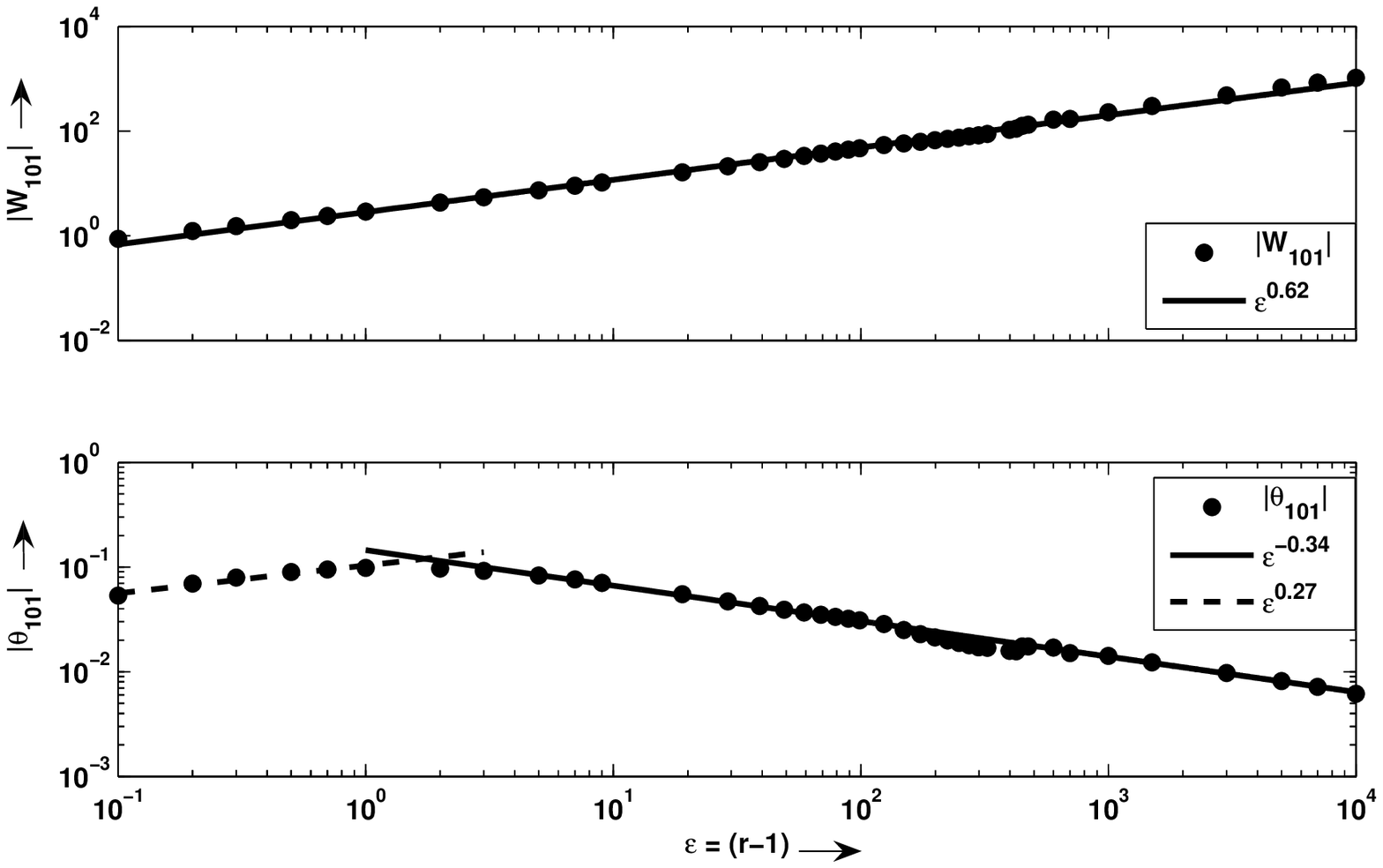}
\end{center}
\caption{}
\label{modes_scal}
\end{figure}

\begin{figure}
\begin{center}
\includegraphics[height=!,width=!]{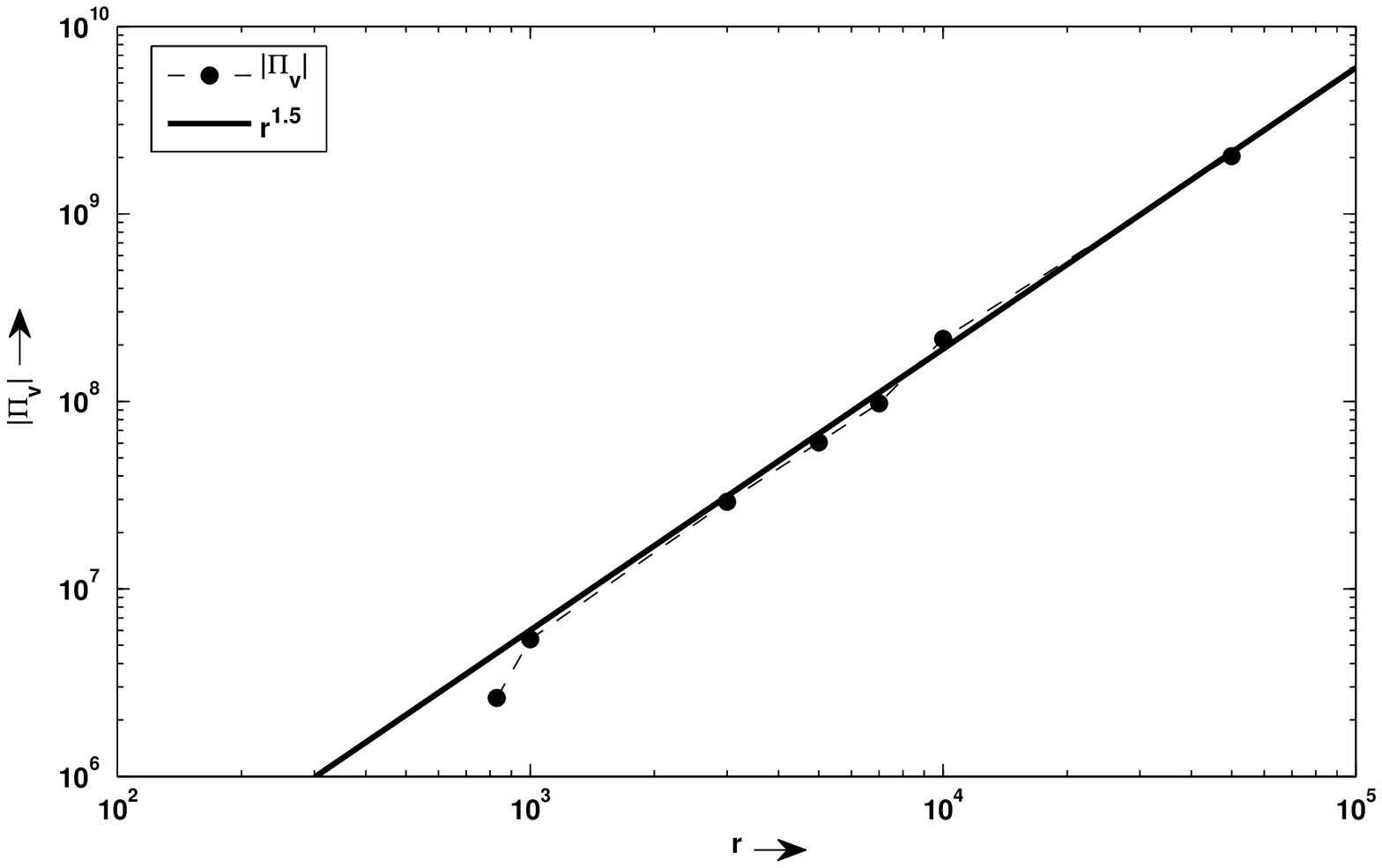}
\end{center}
\caption{}
\label{Piu_scal}
\end{figure}

\begin{figure}
\begin{center}
\includegraphics[height=!,width=!]{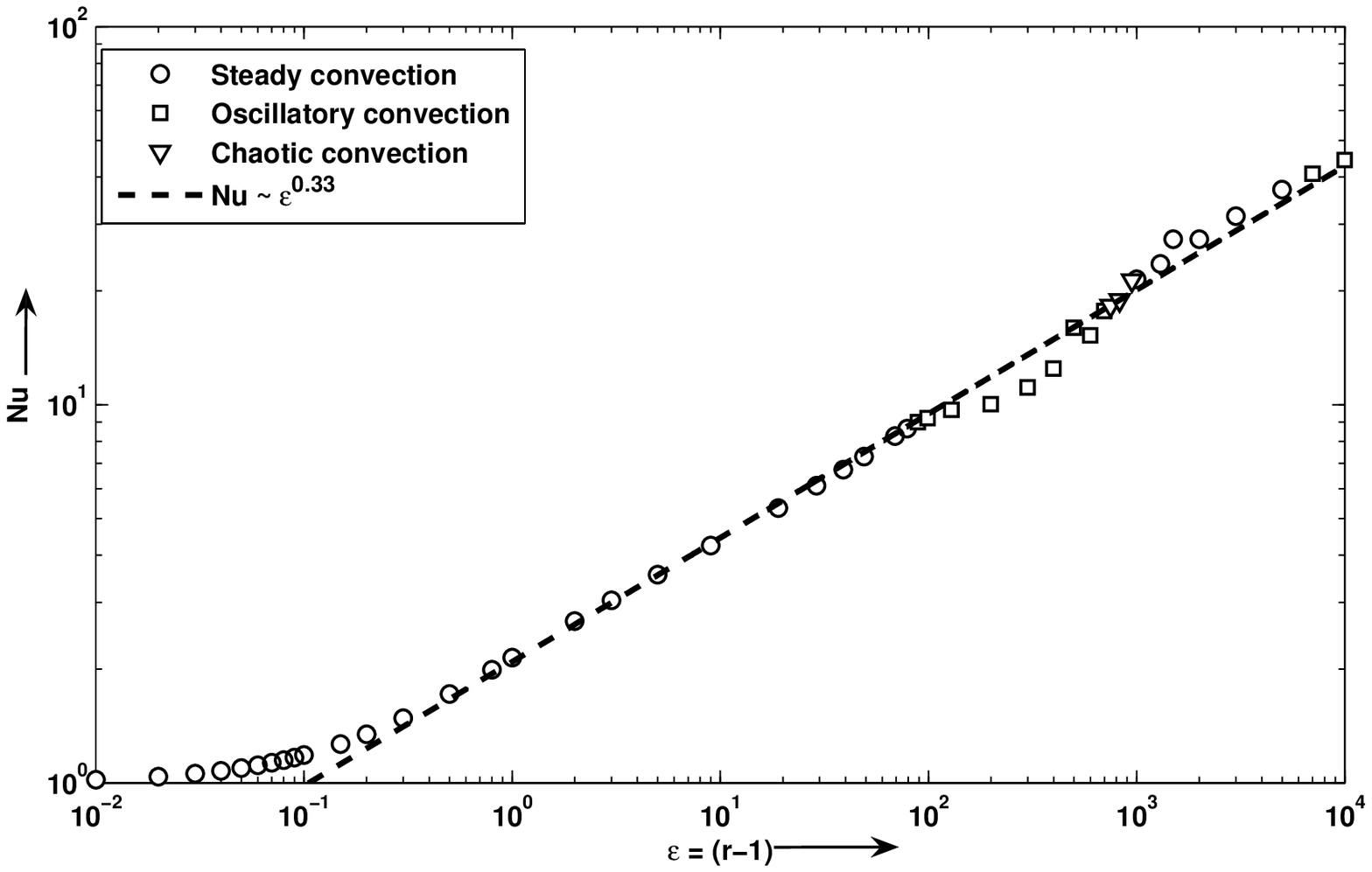}
\end{center}
\caption{}
\label{Nu_scal}
\end{figure}

\noindent
FIG. 2: Phase plots in the $\Im(W_{101})$ - $\Im(\theta_{101})$ plane. The values of $r$ and $P$ are same as in the corresponding plots of figure~\ref{timeseries_w11}.\\

\noindent
FIG. 3: Travelling rolls at $r=830$ ($R= 5.5\times 10^{5}$) and $P = 6.8$. Here red represents hot fluid, and blue represents cold fluid.
The convection rolls travel chaoticlly to the left. The three frames are taken at dimensionless times $t=1.4, 1.43$ and $1.5$ respectively. \\

\noindent
FIG. 4: Kinetic energy spectra $E_v(\kb)$ vs. wavenumber are plotted in (a), (c) and (e) for $r=70$ ($64\times 64$), $140$($128\times128$) and $400$ ($512\times512$) respectively. In (b), (d) and (f), entropy spectra $E_{\theta}(\kb)$ vs. wavenumber are plotted  for the same sequence of $r$. \\

\noindent
FIG. 5: (a) Plots of kinetic energy spectra $E_v(\kb)$ and (b) entropy spectra $E_{\theta}(\kb)$ as a function of $k$ 
at $r=830$. Similar plots for  $r=10000$ (c and d), and for $r=50000$ (e and f). \\

\noindent
FIG. 6: (a) Plots of flux of kinetic energy $\Pi_u$ and (b) entropy $\Pi_{\theta}$ as a function of wavenumber sphere radii at $r=400$. Similar plots for  $r=830$ (c and d), and for  $r=10000$ (e and f). \\

\noindent
FIG. 7: (a) Plots of shell-to-shell energy transfer for $v \rightarrow v$ channel ($T^{vv}_{nm}$) and (b) for $\theta \rightarrow \theta$ 
channel ($T^{\theta\theta}_{nm}$) as a function of $m$ and $n$ at $r=400$. In the figure, red is maximum positive energy transfer, while blue 
is maximum negative energy transfer.  Similar plots for $r=830$ (c and d),  $r=1000$ (e and f), and $r=10000$ (g and h).  \\

\noindent
FIG. 8: Variation of mean $|W_{101}|$ and $|\theta_{101}|$ with $\epsilon=(r-1)$. $|W_{101}|$ 
varies as $\epsilon^{0.62}$.  $|\theta_{101}|$ varies as $\epsilon^{0.267}$ for $\epsilon < 1$ and 
as $\epsilon^{-0.34}$ for $\epsilon>1$. \\

\noindent
FIG. 9: Variation of maximum $|\Pi_v|$ as function of  reduced Rayleigh number $r$. \\

\noindent
FIG. 10: Nusselt number vs. $\epsilon=(r-1)$. The Nusselt number shows a power-law scaling with $\epsilon$. 
The power-law exponent is $0.33 \pm 0.01$. \\

\newpage

\clearpage
\newpage

\clearpage
\newpage

\newpage

\newpage

\newpage

\newpage

\newpage

\clearpage
\newpage

\clearpage
\newpage

\end{document}